\documentclass[aps,prd,twocolumn,amsmath,amssymb,showpacs,floatfix,superscriptaddress,nofootinbib]{revtex4-1}

\usepackage{natbib}
\usepackage{amsmath}
\usepackage{amssymb}
\usepackage{latexsym}
\usepackage{bm}   
\usepackage{graphicx,epstopdf}
\usepackage{float}
\usepackage{epstopdf}
\DeclareGraphicsExtensions{.ps}
\usepackage{dcolumn} 
\usepackage{multirow}
\usepackage{appendix}
\usepackage{footnote}
\usepackage{tabularx,ragged2e,booktabs}

\newcolumntype{C}[1]{>{\Centering}m{#1}}

\usepackage{color}

\DeclareMathAlphabet\mathbfcal{OMS}{cmsy}{b}{n}
\definecolor{darkgreen}{cmyk}{0.85,0.2,1.00,0.35} 
\definecolor{purple}{cmyk}{0.5,1.0,0,0} 
\definecolor{darkblue}{cmyk}{1.0,1.0,0,0}
\usepackage{hyperref}

\hypersetup{
   colorlinks = true,
   allcolors = darkblue}

\newcommand{\AL}{M_L}
\newcommand{\nsims}{n_{\rm sim}}

\def\beq{\begin{equation}}
\def\eeq{\end{equation}}
\def\bea{\begin{eqnarray}}
\def\eea{\end{eqnarray}}
\def\lsim{\mathrel{\raise.3ex\hbox{$<$\kern-.75em\lower1ex\hbox{$\sim$}}}}
\def\gsim{\mathrel{\raise.3ex\hbox{$>$\kern-.75em\lower1ex\hbox{$\sim$}}}}
\def\wigner#1#2#3#4#5#6{ \left( \begin{array}{ccc} #1 & #3 & #5
\\ #2 & #4 & #6 \\ \end{array} \right)}

\newcommand{\bn}{\hat{\bm n}}

\begin{document}
	
\title{Lensing Bias to CMB Measurements of Compensated Isocurvature Perturbations}

\author{Chen He Heinrich}\email{chenhe@uchicago.edu}
\affiliation{Department of Physics, University of Chicago, Chicago, IL 60637}
\affiliation{Kavli Institute for Cosmological Physics, Enrico Fermi Institute, Chicago, IL 60637}

\author{Daniel Grin}
\affiliation{Kavli Institute for Cosmological Physics, Enrico Fermi Institute, Chicago, IL 60637}
\affiliation{Department of Astronomy and Astrophysics, University of Chicago, Chicago, IL 60637} 

\author{Wayne Hu}
\affiliation{Kavli Institute for Cosmological Physics, Enrico Fermi Institute, Chicago, IL 60637}
\affiliation{Department of Astronomy and Astrophysics, University of Chicago, Chicago, IL 60637} 

\begin{abstract}
Compensated isocurvature perturbations (CIPs) are modes in which the baryon and dark matter density fluctuations cancel. They arise in the curvaton scenario as well as some models of baryogenesis.
While they leave no  observable effects on the cosmic microwave background (CMB) at linear order, they do spatially modulate two-point CMB statistics and can be reconstructed in a manner similar to gravitational
lensing.    Due to the similarity between the effects of CMB lensing and CIPs, lensing contributes nearly Gaussian random noise to the CIP estimator that approximately doubles the reconstruction noise power.  Additionally, the cross correlation between lensing and the integrated Sachs-Wolfe (ISW) effect generates a correlation between the CIP estimator and the
temperature field even in the absence of a correlated CIP signal.  
For cosmic-variance limited temperature measurements out to multipoles $l \leq  2500$,  subtracting a fixed lensing bias  degrades the detection threshold for CIPs by a factor of $1.3$, whether or not they are correlated with the adiabatic mode.
\end{abstract}
\maketitle

\section{Introduction}

Measurements of the cosmic microwave background (CMB) are consistent with  primordial curvature fluctuations that are nearly scale-invariant, Gaussian, and adiabatic \cite{Ade:2013uln,Ade:2015lrj,Ade:2015ava}. 
They lend empirical support to the idea that these perturbations were produced during inflation, an epoch of accelerated cosmic expansion, as quantum fluctuations of a single field (the \textit{inflaton}) \cite{Guth:1982ec,Bardeen:1983qw,Mukhanov:1990me,Kofman:1994rk,Amin:2014eta}.

It is still possible, however, that two different fields drive inflationary expansion and seed primordial fluctuations, leaving small but observable entropy or isocurvature fluctuations \cite{Bond:1984fp,Kodama:1986fg,Kodama:1986ud,Hu:1994jd,Moodley:2004nz,Bean:2006qz}.  For example, in the
\emph{curvaton} model 
a spectator field during inflation comes to dominate the density and hence produces curvature
fluctuation after inflation.  If the curvaton produces baryon number, lepton number, or cold dark matter (CDM) as well, it could lead to a mixture of curvature and isocurvature fluctuations \cite{Mollerach:1989hu,Mukhanov:1990me,Moroi:2001ct,Lyth:2001nq,Lyth:2002my,Postma:2002et,Kasuya:2003va,Ikegami:2004ve,Mazumdar:2004qv,Allahverdi:2006dr,Papantonopoulos:2006xi,Enqvist:2009zf,Mazumdar:2010sa,Mazumdar:2011xe}.  
Given their common source in the curvaton field fluctuations, these isocurvature and curvature modes are 
typically correlated, which affects their observability \cite{Lyth:2001nq,Lyth:2002my}.

Both correlated and uncorrelated isocurvature fluctuations between the non-relativistic matter (baryons and CDM) and photons are highly constrained by CMB observations to be much smaller than curvature
fluctuations \cite{Bucher:2000hy,Valiviita:2003ty,KurkiSuonio:2004mn,Beltran:2005xd,MacTavish:2005yk,Keskitalo:2006qv,Komatsu:2008hk,Valiviita:2009bp,Komatsu:2010fb,Mangilli:2010ut,Kasanda:2011np,Hinshaw:2012aka,Valiviita:2012ub,Kawasaki:2014fwa,Ade:2015lrj,Ade:2015ava}. 
In the curvaton scenario, these observations {impose} constraints {to} curvaton decay scenarios \cite{Lyth:2003ip,Gordon:2002gv,Beltran:2004uv,Beltran:2008aa,Gordon:2009wx,DiValentino:2011sv,DiValentino:2014eea,Ade:2015ava,Smith:2015bln}.  Nonetheless, there is an additional isocurvature mode, called a compensated isocurvature perturbation (CIP), that is allowed to be of order the curvature fluctuation or
larger so long as the
decay parameters  lie in a range allowed by the matter isocurvature constraints
\cite{Lewis:2002nc,cambnotes,Holder:2009gd,Gordon:2009wx,Ade:2015ava}.  These
 modes entirely evade  linear theory constraints from the CMB  because the CDM and baryon
isocurvature fluctuations are compensated so as to
 produce no early-time gravitational potential or radiation pressure perturbation \cite{Lewis:2002nc,cambnotes,Grin:2011nk}. 
In addition to the curvaton model, CIPs could be produced in some models of baryogenesis \cite{DeSimone:2016ofp}.

At higher order, the CIP  modulation of the baryon density leads to spatial fluctuations in the diffusion-damping scale and acoustic horizon of the baryon-photon plasma \cite{Grin:2011nk,Grin:2011tf,Grin:2013uya}. As a result, the CIP field  can be reconstructed using the induced off-diagonal correlations between different multipole moments in CMB temperature and polarization maps.   CIPs also induce a second order change in CMB power spectra \cite{Munoz:2015fdv}.   Although the physics is different, these effects  are very similar to those encountered in the weak gravitational lensing of the CMB \cite{Grin:2011nk,Grin:2011tf}. 

Both effects have been used to limit the amplitude of CIP modes.   
{Using WMAP data, limits from direct reconstruction were imposed in Ref.~\cite{Grin:2013uya}, while limits from CMB power spectra (using \emph{Planck} data) were imposed in Ref.~\cite{Munoz:2015fdv}}.  The latter work showed that the existence of CIPs could even reduce internal tensions in CMB data sets (see, for example \cite{Ade:2015lrj,DiValentino:2015bja,Addison:2015wyg})
but only if the CIP fluctuations are  orders of magnitude larger than the curvature fluctuations.
   
{Future experiments (such as CMB-S4 \cite{Abazajian:2013oma}) that approach the cosmic-variance limit for measurements of all CMB fields out to arcminute scales in principle can provide a detection of correlated CIP modes through reconstruction, with magnitude of
order 10 times the curvature fluctuations, and hence test curvaton decay scenarios
\cite{He:2015msa}}.  {The similarity between CIP reconstruction and gravitational lens
reconstruction from quadratic combinations of CMB fields \cite{Zaldarriaga:1998te,Hu:2001fa,Okamoto:2003zw}, however, suggests that  CIP estimators could  be biased in the presence of CMB lensing.}   Past work has estimated this bias and showed that it does not alter the upper limits to CIPs from WMAP \cite{Grin:2013uya}. Here we explore this issue further, and show that lensing can significantly degrade
the CIP detection threshold from CMB temperature reconstruction for future experiments. 

We begin in Sec.~\ref{sec:background} by summarizing curvaton-model predictions for correlated CIPs, CIP reconstruction techniques, and the origin of lensing bias to CIP measurements. We describe the technique used to simulate lensing and CIP reconstruction as well as  explore the bias lensing produces in the CIP auto and cross-temperature spectra in Sec.~\ref{sec:simulations}.  We then perform a Fisher-matrix analysis to evaluate the impact of CMB lensing on the sensitivity of a cosmic-variance limited experiment to correlated and uncorrelated CIPs in Sec.~\ref{sec:results}. We discuss the implications and conclude in Sec~\ref{sec:conclusions}.

\section{Compensated Isocurvature Perturbations}
\label{sec:background}

In this section we briefly review the origin of and observable imprint in the
CMB of compensated isocurvature perturbations.   We refer the reader to
Refs.~\cite{Grin:2011tf,He:2015msa} for more details.

\subsection{Modulated Observables}

The primordial perturbations of the early Universe can be decomposed into 
curvature fluctuations on constant density slicing $\zeta$, and entropy fluctuations in
the relative number density fluctuations of the various species $n_i$ with respect to the
photons
	\begin{equation}
		S_{i\gamma}=\frac{\delta n_{i}}{n_{i}}-\frac{\delta n_{\gamma}}{n_{\gamma}}.
	\end{equation} 
Here $i \in 	\left\{b,c,\nu,\gamma\right\}$ with $b$ for baryons, $c$ for cold dark matter (CDM), $\nu$ for neutrinos, and $\gamma$ for photons.
The compensated isocurvature mode $\Delta$ is a special combination of entropy fluctuations for which
the baryon-photon number density fluctuates but is exactly compensated by the CDM
in its energy density perturbations
	\begin{eqnarray}
	S_{b\gamma} = \Delta ,\quad	S_{c\gamma} = -\frac{\rho_b}{\rho_c} \Delta, \quad
	S_{\nu\gamma} = 0.
	\end{eqnarray}
To linear order in the fluctuations, there are no observable effects of this mode in the CMB.
At higher order, the curvature fluctuations and the acoustic waves they generate propagate
in a medium with spatially varying baryon to photon and baryon to CDM ratios.   For CIP modes that are
larger than the sound horizon at recombination, subhorizon modes behave as if they
were in a separate universe with perturbed cosmological parameters \cite{Grin:2011tf}
	\begin{align} 
	\delta \Omega_b = \Omega_b \Delta, \quad \delta \Omega_c = - \Omega_b\Delta.
	\label{eqn:separate}
	\end{align}
In this limit, we can take the usual calculation for the CMB temperature power spectrum given a curvature power spectrum $P_{\zeta\zeta}$
	\begin{align}
			C_l^{\tilde T\tilde T} = \frac{2}{\pi}\int k^{2}dk T_{l}^{\tilde T}(k) T_l^{\tilde T}(k)P_{\zeta\zeta}(k),
	\label{eqn:ClTT}
	\end{align}
and note that its dependence on the baryon and CDM background densities come
solely through the radiation transfer functions  {$T_l^{\tilde{T}}(k)$}.   For a small $\Delta$,
 we can Taylor expand the power spectrum to extract its sensitivity to CIPs through the derivative
 \begin{align}
			C_l^{\tilde T,d\tilde T} = \frac{2}{\pi}\int k^{2}dk T_{l}^{\tilde T}(k) \frac{d T_l^{\tilde T}}{d\Delta}(k)P_{\zeta\zeta}(k).
			\label{eqn:Clderiv}
	\end{align}
	
Since this separate universe approximation involves a spatial modulation at recombination, the radiation transfer functions and $C_l^{\tilde T\tilde T}$
represent the power spectrum in the absence of gravitational lensing.    Formally, they should also omit post recombination effects from reionization and the
integrated Sachs-Wolfe effect but since we are interested in the change in the subhorizon modes we simply employ the usual radiation transfer functions.

The implied position-dependent power spectrum when considered on the whole sky represents a squeezed bispectrum in the CMB where 
the superhorizon mode is taken to be much larger than the subhorizon modes.   
   Quadratic
combinations of subhorizon modes can be used to reconstruct the superhorizon CIP modes in the same manner as in CMB gravitational lens reconstruction \cite{Grin:2011tf}
as long as the CIP mode is larger than the sound horizon in projection at recombination, i.e. for multipoles $l \lesssim 100$ \cite{He:2015msa}.  We explicitly construct this estimator of the CIP field $\Delta$ in {Sec.~\ref{sec:simulations}}.
To the extent that the CIP and lensing modulations share the same structure, one will contaminate the other.
\subsection{Curvaton CIPs}
One scenario in which CIPs arise is the curvaton scenario, where a a spectator scalar field during inflation later creates  curvature perturbations. This field (the \emph{curvaton}). It then decays into other particles, and depending on whether it generates baryon number or CDM in doing so, isocurvature perturbations can arise and are correlated with curvature fluctuations. In particular, the presence of
CIP models distinguishes between possible decay scenarios of the curvaton model \cite{He:2015msa}.

If the curvaton
gives rise to all of the curvature perturbation $\zeta$, then any resulting CIPs would be fully correlated with it
    \begin{equation}
   \Delta = A \zeta,
    \end{equation}
where $A$ determines the amplitude of the CIP.
There are two decay scenarios that are particularly interesting because of their relatively large
CIP amplitude: $A \approx 3\Omega_c/\Omega_b$ if baryon number is produced by the curvaton decay and CDM before curvaton decay, while $A= -3$ if CDM is produced by curvaton decay, and baryon number is produced before curvaton decay.

We can exploit the correlated nature of curvaton CIPs by measuring the cross spectra \cite{He:2015msa}
	\begin{align}
			C_l^{XY} =  \frac{2}{\pi}\int k^{2}dk T_{l}^X(k)  T_l^Y(k)
			P_{\zeta\zeta}(k),
	\label{eqn:Cl}
	\end{align}
with $X,Y \in \tilde T,\Delta$ where the CIP transfer function
\begin{align}
T_l^\Delta(k) = A j_l(kD_*)
\end{align}
 represents a simple projection of $\Delta$ onto the shell at the distance $D_*$ to recombination in a spatially
flat universe. 
For a {low signal-to-noise} reconstruction of CIP modes, the cross correlation enhances the ability of the CMB to detect
the CIP modes (see Ref. \cite{He:2015msa} and Tab.~\ref{tab:2sigma} below).

Note that in practice we can only cross correlate the lensed CMB field to determine $C_l^{T\Delta}$ though
on the relevant $l \lesssim 200$ scales the difference between the lensed and {unlensed CMB} is negligible.  On the other hand, we shall see
that the ISW-lensing cross correlation \cite{Goldberg:1999xm,Smith:2006ud,Lewis:2011fk,Kim:2013nea,Ade:2015dva} provides a non-zero signal for this measurement even in the absence of a true CIP mode.
\section{Simulations}
\label{sec:simulations}
In this section, we simulate CIP reconstruction from CMB temperature sky maps 
to assess  its noise properties with and without non-Gaussian contributions from CMB
lensing.  
We take a flat $\Lambda$CDM cosmology consistent with the Planck 2015 results \cite{Ade:2015xua}\footnote{Specifically, we use results obtained with the TT, TE, EE + lowP likelihood.} with baryon density $\Omega_b h^2$ = 0.02225, cold dark matter density $\Omega_c h^2$ = 0.1198, Hubble constant $h = 0.6727$, scalar amplitude $A_s = 2.207\times10^{-9}$, spectral index $n_s = 0.9645$,  reionization optical depth $\tau = 0.079$, neutrino mass of a single species contributing to the total $\Omega_m$ = 0.3156, and $T_{\rm cmb}$ = 2.726K. 
The lensing simulations are performed using  CAMB,\footnote{CAMB: \url{http://camb.info}} 
LensPix,\footnote{LensPix: \url{http://cosmologist.info/lenspix/}} and HEALPix\footnote{HEALPix: \url{http://healpix.sourceforge.net}} and 
CIP reconstructions employ a modified version of LensPix that we describe below.

\begin{figure}
          \includegraphics[width=0.8\columnwidth]{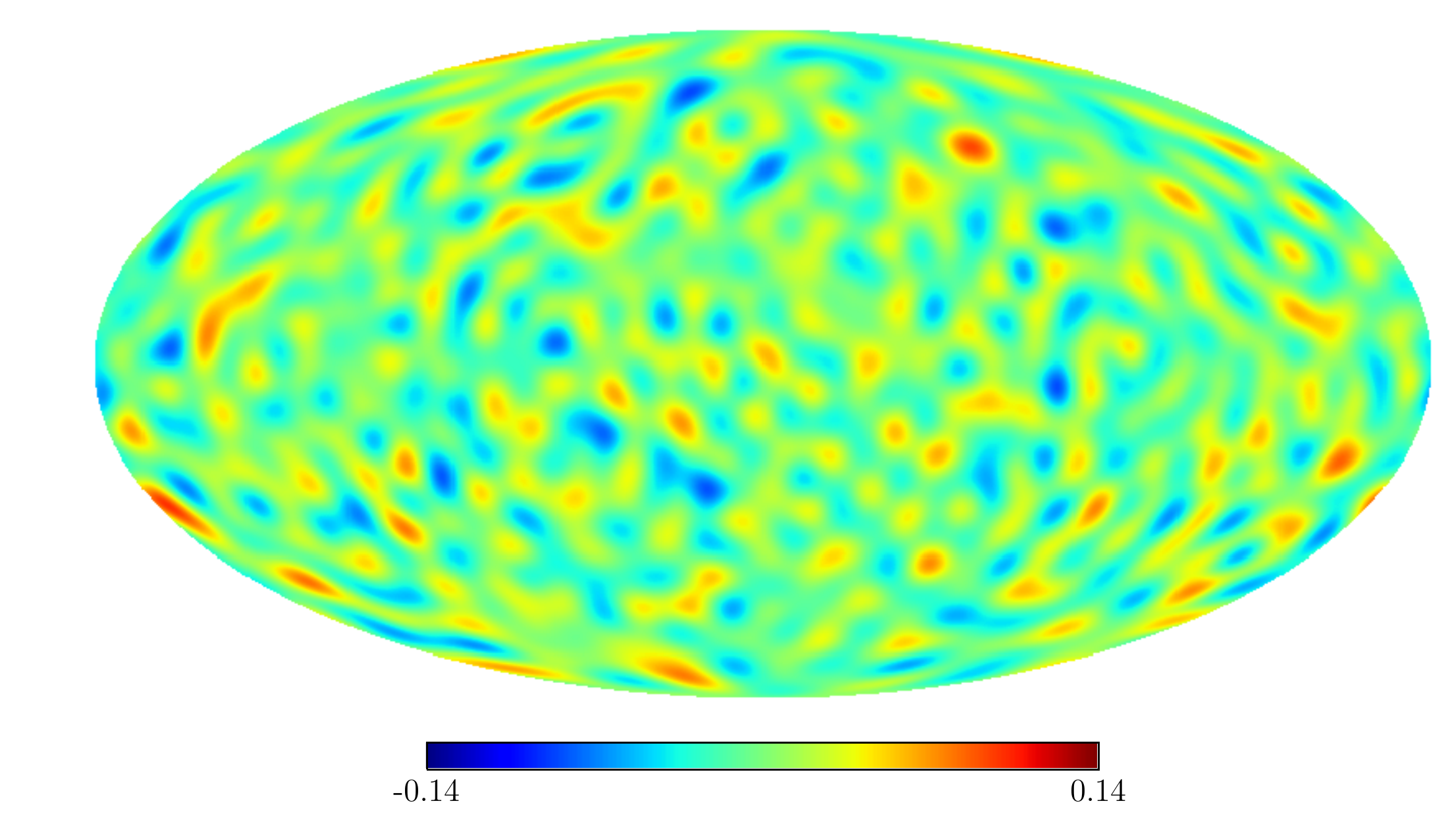}      
          \includegraphics[width=0.8\columnwidth]{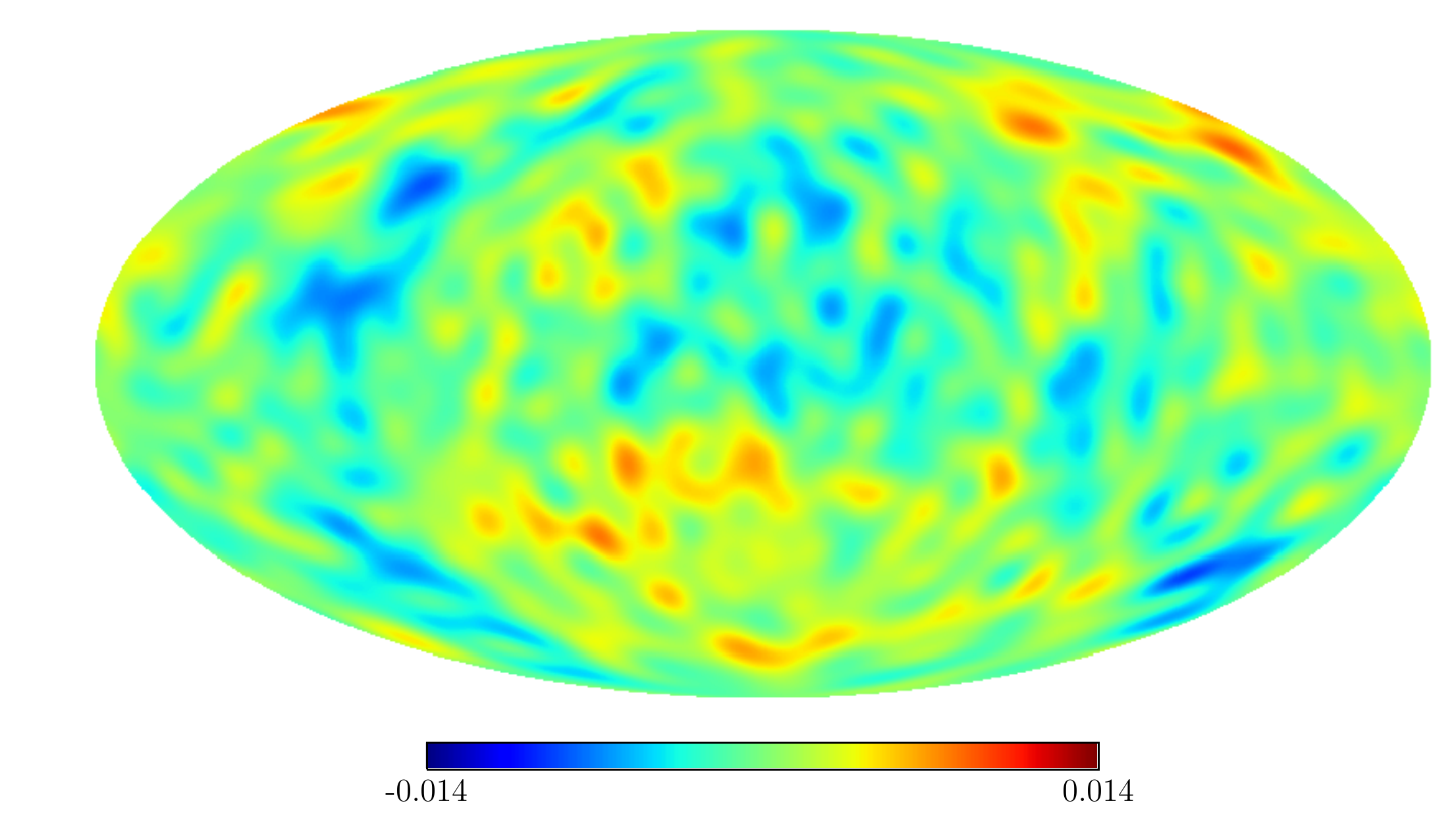}
         \caption{Top: Realization of CIP reconstruction noise assuming a Gaussian CMB temperature field.  Bottom:
         target CIP signal with $A=30$ which is nearly scale-invariant shown with a factor of 10 smaller range than the nearly white noise.  
         Maps have been low pass filtered to {$l \leq 30$}           to highlight modes where in the
         curvaton scenario, the temperature-CIP cross correlation can be used to detect the signal. }
         \label{fig:mapgaussian}
          \end{figure}
          
\subsection{CIP Reconstruction}

To test the reconstruction pipeline, we begin with the case for which the CIP quadratic estimator was designed \cite{Grin:2011tf}, an otherwise Gaussian random CMB temperature field. 
In fact, we take the amplitude of the CIP signal to zero to simulate the noise properties
of the estimator and in this case the CMB temperature field is  completely 
Gaussian by construction.

Using HEALPix, we draw $\nsims = 4000$ Gaussian random realizations
of the dimensionless temperature fluctuation field $\hat T_{lm}$ from the CMB temperature spectrum $C_l^{TT}$. 
  Note that this
is the power spectrum of the lensed CMB but here the $\hat T_{lm}$ do not contain the non-Gaussian
correlations of the properly lensed CMB sky.  We do this so that we can isolate the
non-Gaussian aspects of lensing below.
We take  $N_{\rm side} = 2048$ and $l_{\rm max} = 3900$; we have verified that these values yield sufficient accuracy to use modes out
to $l=2500$ in the estimator.

From each realization we construct the inverse variance and derivatively filtered fields
	\begin{eqnarray}
		\hat T_V(\bn) &=& \sum_{l = 2}^{2500} \sum_{m=-l}^{l}\frac{1}{C_l^{\tilde T\tilde T}} \hat T_{lm}  Y_{lm} (\bn) ,	\nonumber \\
		\hat T_D(\bn) &=& \sum_{l = 2}^{2500} \sum_{m=-l}^{l} \frac{C_l^{\tilde T,d\tilde T} }{C_l^{\tilde T\tilde T}}	\hat T_{lm} Y_{lm} (\bn) ,
		\label{eqn:filteredfields}
	\end{eqnarray}
where $C_l^{\tilde T \tilde T}$ is the power spectrum of the unlensed CMB.  These filtered fields
are then combined to form the  minimum variance quadratic
estimator for  cosmic-variance limited temperature measurements out to $l=2500$  \cite{Grin:2011tf,Grin:2013uya}
	\begin{equation}
		\hat{\Delta}_{LM} = \AL \int d\bn Y_{LM}^{*} (\bn) {\hat T}_V(\bn) {\hat T}_D(\bn).
		\label{eqn:estimator}
	\end{equation} 
Here the normalization\footnote{We correct a small $(L+1)M_L \rightarrow L M_L$ error in the numerical results on reconstruction noise and parameter errors obtained from Ref.~\cite{He:2015msa}, Fig. 2.}
        \begin{align}
    		\frac{1}{\AL} & = \sum_{l ,l'= 2}^{2500}
		 \frac{(2l+1)(2l'+1)}{4\pi}\wigner{l}{0}{L}{0}{l'}{0} ^2\nonumber\\
		&\quad  \times 
		\frac{ \big[C^{\tilde T,d\tilde T}_{l'} + C^{\tilde T,d\tilde T}_{l}\big]^2}{2 C_{l}^{\tilde T\tilde T}C_{l'}^{\tilde T\tilde T}}
		\label{eqn:normalization}	
    \end{align}    
would return an unbiased estimator of the CIP field in the separate universe approximation and in
the absence of lensing.  Specifically, given that our realizations lack a true CIP signal, the 
average over the ensemble of temperature realizations
\begin{equation}
\langle \hat{\Delta}_{LM}  \rangle = 0.
\end{equation}
In Fig.~\ref{fig:mapgaussian}, we show a single realization of the zero mean noise in the estimator.  
The noise of the estimator is nearly white.
For reference we compare it to a realization of a true CIP signal with $A=30$ which is nearly scale-invariant
and hence has relatively more power on large scales.
We have also tested this estimator against CMB temperature maps with a true CIP
signal following the procedure of Ref.~\cite{Grin:2013uya} and find that the reconstruction is unbiased in the absence of lensing-induced off-diagonal correlations.

\begin{figure}
          \includegraphics[width=\columnwidth]{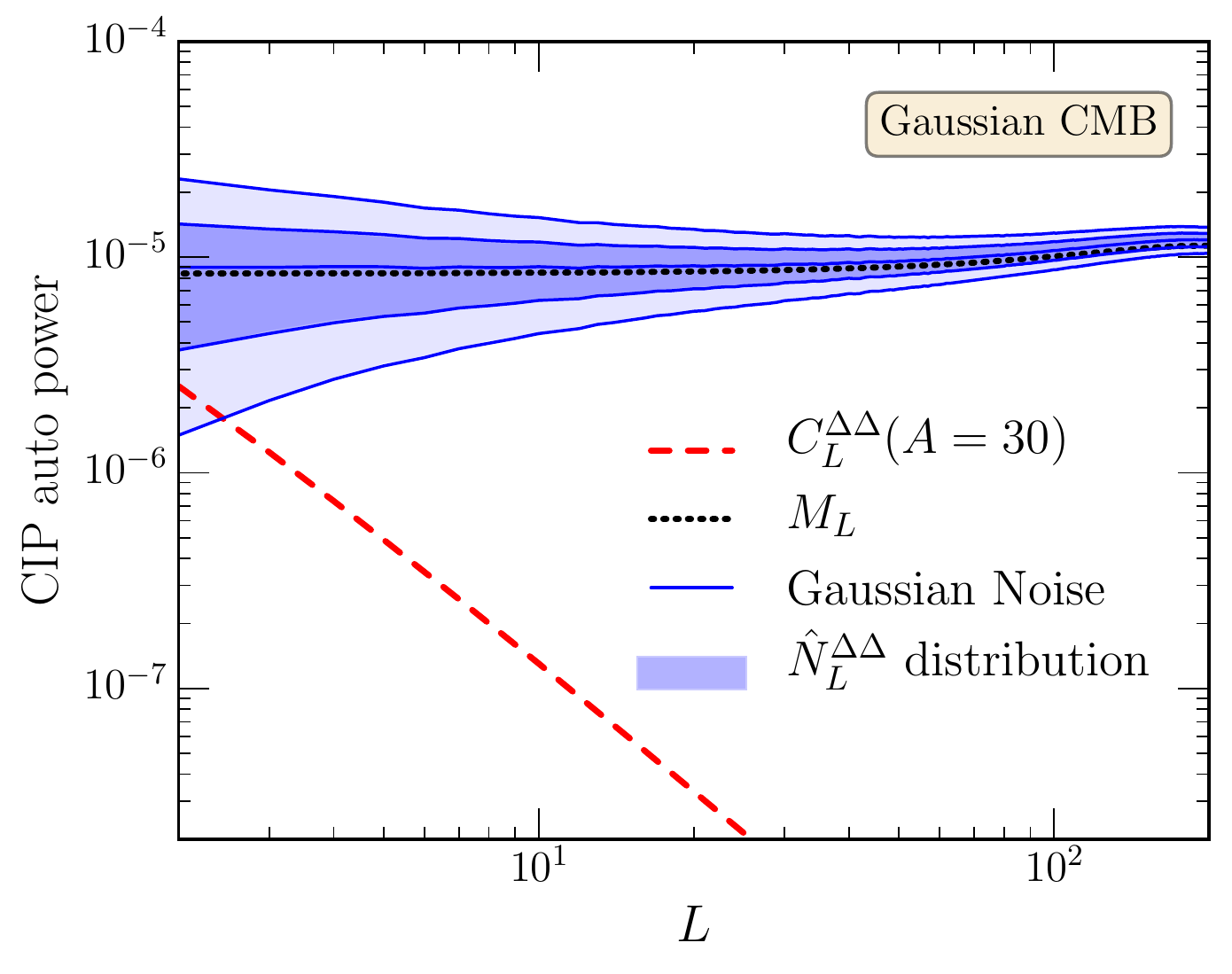}
          \caption{
          CIP estimator noise power $\hat{N}_{L}^{\Delta\Delta}$ assuming Gaussian CMB maps.
          Shown are the mean, 68\% and 95\% confidence bands (shaded) of 4000 realizations of the estimator in
          the absence of a CIP signal.   The mean matches closely the theoretical expectation 
          $\AL$ (dotted line) from Eq.~(\ref{eqn:normalization}).   The confidence bands match the $\chi^2$ expectation of
          Gaussian noise given the mean (solid lines).  For reference we show a true CIP signal with $A=30$ (dashed line).
}
          \label{fig:DDgaussian}
\end{figure}

\begin{figure}
          \includegraphics[width=\columnwidth]{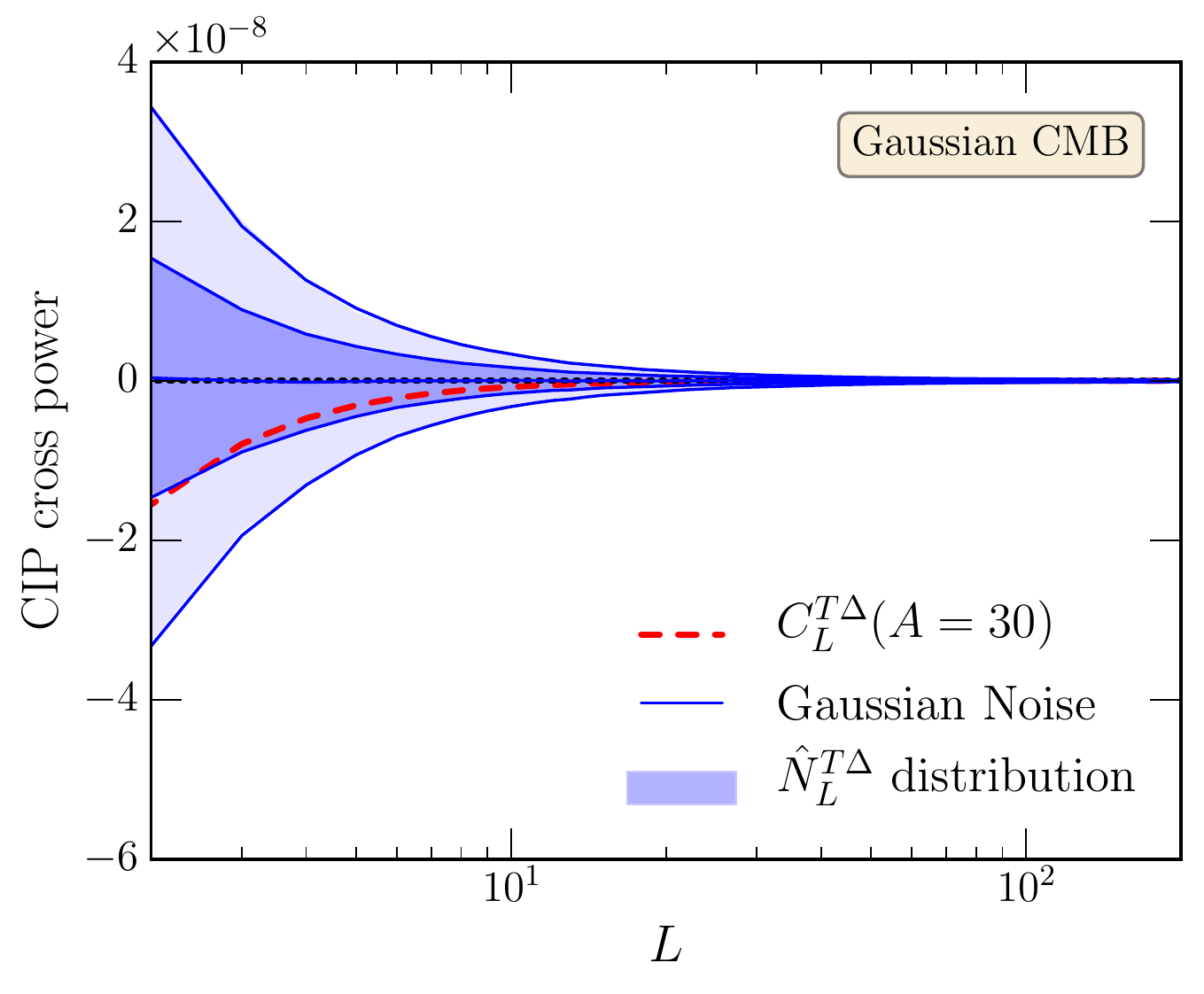}
          \caption{
          CIP noise cross power $\hat{N}_{L}^{T\Delta}$ assuming Gaussian CMB maps.
          Shown are the mean, 68\% and 95\% confidence bands (shaded) of 4000 realizations of the estimator in
          the absence of a CIP signal.   The mean matches closely the theoretical expectation $N_L^{T\Delta}=0$ (dotted line).      	  The confidence bands match the Wishart expectation of
          Gaussian noise  (solid lines, see text).     For reference we show a true correlated CIP signal with $A=30$ (dashed line).
}
          \label{fig:TDgaussian}.
\end{figure}

From each realization we construct the all-sky estimator of the auto and cross power spectra of
the CIP noise
\begin{equation}
\hat{N}_L^{XY} =  \frac{1}{2L+1} \sum_{M} \hat {X}_{LM}^{*} \hat{ Y}_{LM},
\label{eqn:noiserealization}
\end{equation}
for $X,Y \in T,\Delta$.   Averaging the estimators  over the ensemble of realizations 
yields the noise power spectra
\begin{equation}
N_L^{XY} = \langle \hat{N}_L^{XY}  \rangle.
\label{eqn:noiseaverage}
\end{equation}
We compare these simulated spectra to the theoretical expectation 
$N_L^{\Delta\Delta}=\AL$
and $N_L^{T\Delta}=0$
in Figs.~\ref{fig:DDgaussian} and \ref{fig:TDgaussian}  and find good agreement.
Given the quadratic nature of the $\Delta$ estimator, $N_L^{T\Delta}$ is proportional
to the temperature bispectrum which vanishes for a Gaussian temperature field.
Note that this agreement also tests the slight mismatch in normalization versus noise due to the use of
$C_l^{TT}$ for the realizations whereas the filters are built out of the unlensed $C_l^{\tilde T\tilde T}$.
We have also tested that $N_L^{TT} = C_L^{TT}$ as expected.

Using the distribution of realizations $\hat N_L^{XY}$ we can also test the approximation
that the cosmic variance of the quadratic combinations of the temperature field  produces nearly Gaussian random noise in the CIP estimator.  While the product of Gaussian variates
is not Gaussian distributed, the estimator is formed out of many combinations of temperature
multipoles.    The central limit theorem implies that the noise in the estimator should be 
much closer to Gaussian than the product of any individual pair.

If the  noise in the CIP estimator and the CMB temperature field are themselves 
Gaussian distributed,
the power spectra estimators $\hat{N}_L^{XY}$ should be distributed according to a rank 2 Wishart distribution of $2L+1$ degrees of freedom
\begin{equation}
P(\hat{\bf N}_L| {\bf N}_L ) = W_2\left( \frac{{\bf N}_L}{2L+1} , 2L+1 \right).
\end{equation}
The Wishart distribution gives the joint probability density of the set of
power spectra $\hat{N}_L^{XY}$
given $N_L^{XY}$ which form $2\times 2$ symmetric matrices $\hat {\bf N}_L$ and 
${\bf N}_L$ with elements $X,Y \in \{T,\Delta\}$.
To obtain the marginal probability distribution for a single spectrum e.g.\ $\hat N_L^{T\Delta}$  one
integrates the joint distribution over the other spectra (see e.g. Ref.~\cite{Percival:2006ss}).
The Wishart scale matrix ${\bf N}_L/(2L+1)$ reflects the fact that $\hat{\bf N}_L$ is defined as the average over the $2L+1$ $M$-modes of the
products of the Gaussian fields rather than the sum.    For example, rescaling the statistic
\begin{equation}
\frac{2L+1}{ N_L^{XX}} \hat N_L^{XX} = \sum_{M=-L}^{L}\frac{ \hat X_{LM}^*\hat X_{LM}} {N_L^{XX}}
\label{eqn:chi2}
\end{equation}
brings its marginal distribution to  a $\chi^2$ of $2L+1$ degrees of freedom. 

In Figs.~\ref{fig:DDgaussian} and \ref{fig:TDgaussian}, we compare the $68\%$ and
$95\%$ confidence regions of the $\hat{N}_L^{\Delta\Delta}$ and $\hat{N}_L^{T\Delta}$
distributions to the Gaussian expectations of the marginal distributions described above.  Again we find good agreement, indicating
that the noise of the CIP estimator built from quadratic combinations of 
a Gaussian CMB temperature field is itself nearly Gaussian.

\begin{figure}
          \includegraphics[width=0.8\columnwidth]{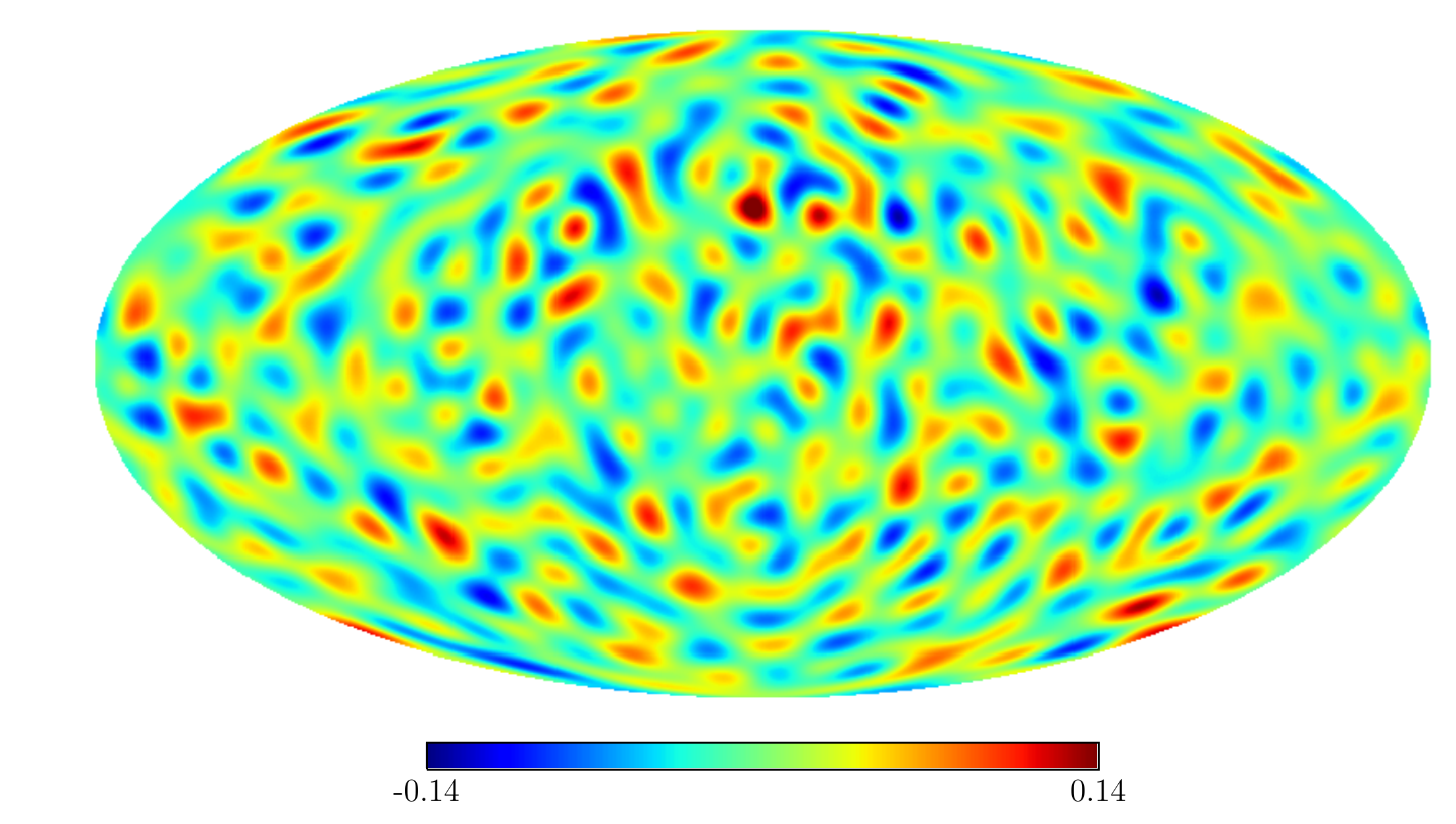}
          \includegraphics[width=0.8\columnwidth]{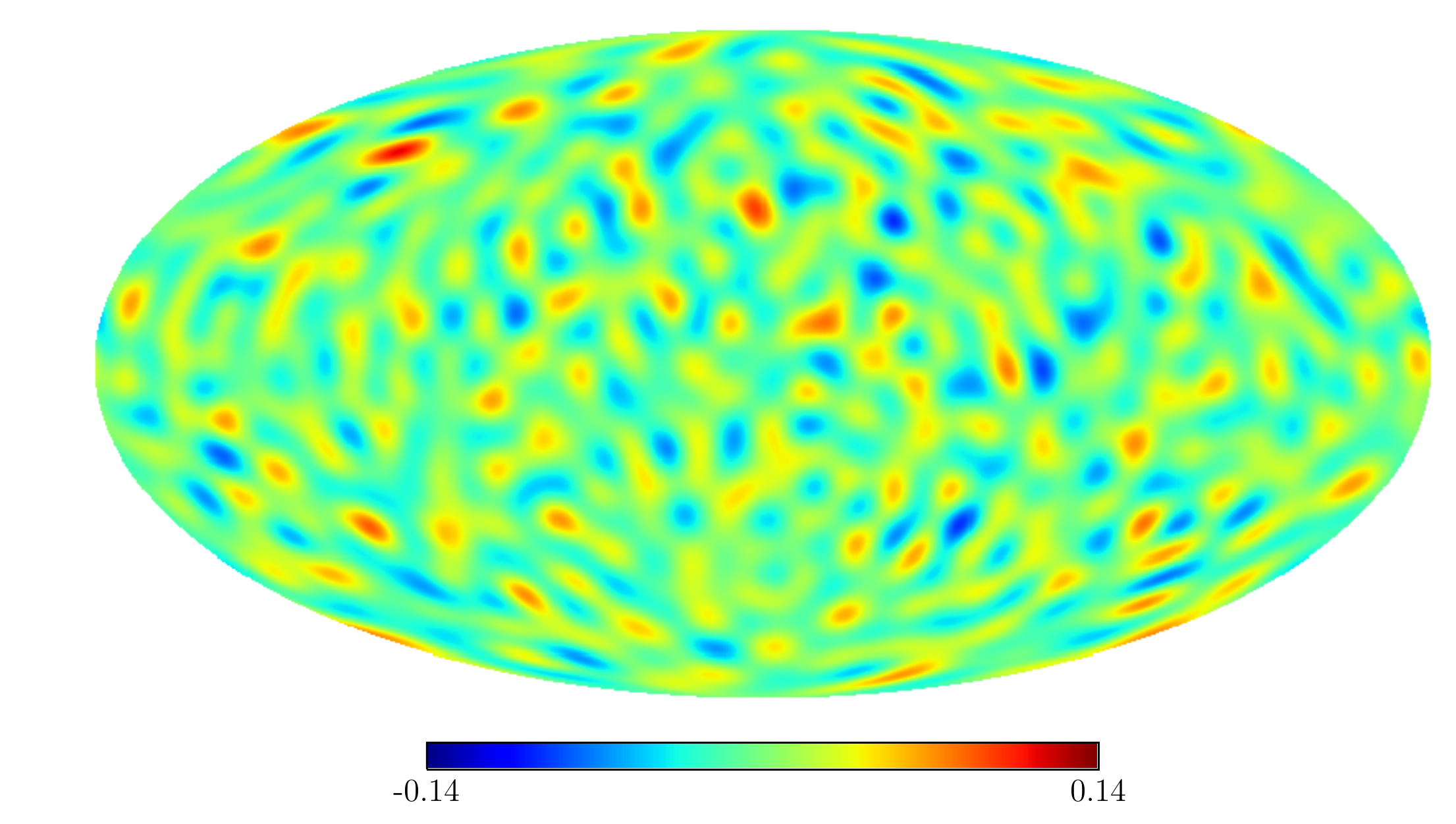}
          \caption{
          Top: Realization of CIP reconstruction noise using a lensed CMB temperature field.  The
          noise is significantly larger than the Gaussian CMB case of Fig.~\ref{fig:mapgaussian}.  Bottom:  CIP reconstruction averaged over 40
          realizations of the CMB sky at recombination lensed by the same potential.   CIP reconstruction is
          biased by the statistical anisotropy of lensing contributing noise that is comparable in amplitude and spectrum
          to the Gaussian case. }
         \label{fig:maplensed}
\end{figure}

\subsection{Lensing Noise}
\label{sec:lensingsim}

Next we test the reconstruction pipeline with the properly lensed and hence non-Gaussian CMB.    In this case we instead
draw $4000$ Gaussian random realizations $\tilde{T}_{lm}$  from the unlensed CMB power spectrum
$C_l^{\tilde T\tilde T}$ and 4000 lensing potentials $\phi_{lm}$ from $C_l^{\phi\phi}$ with correlations  consistent with the $C_l^{\tilde T \phi}$ ISW-lensing cross power  \cite{Smith:2006ud,Lewis:2011fk,Kim:2013nea,Ade:2015dva} as supplied by CAMB.

The angular positions of pixels in the
unlensed map are then remapped into the lensed map according to 
the  gradient of the lensing potential \cite{Zaldarriaga:1998te,Hu:2001fa,Okamoto:2003zw}
 \begin{equation}
 \hat T(\bn) = \tilde T(\bn+\nabla \phi) 
 \end{equation}
 using LensPix.   As in the separate universe approximation for CIPs, a large scale lens acts like a position
 dependent modulation of the CMB and so produces similar effects, in particular an off-diagonal
  two-point
 correlation of multipoles moments which biases CIP reconstruction.

\begin{figure}
          \includegraphics[width=\columnwidth]{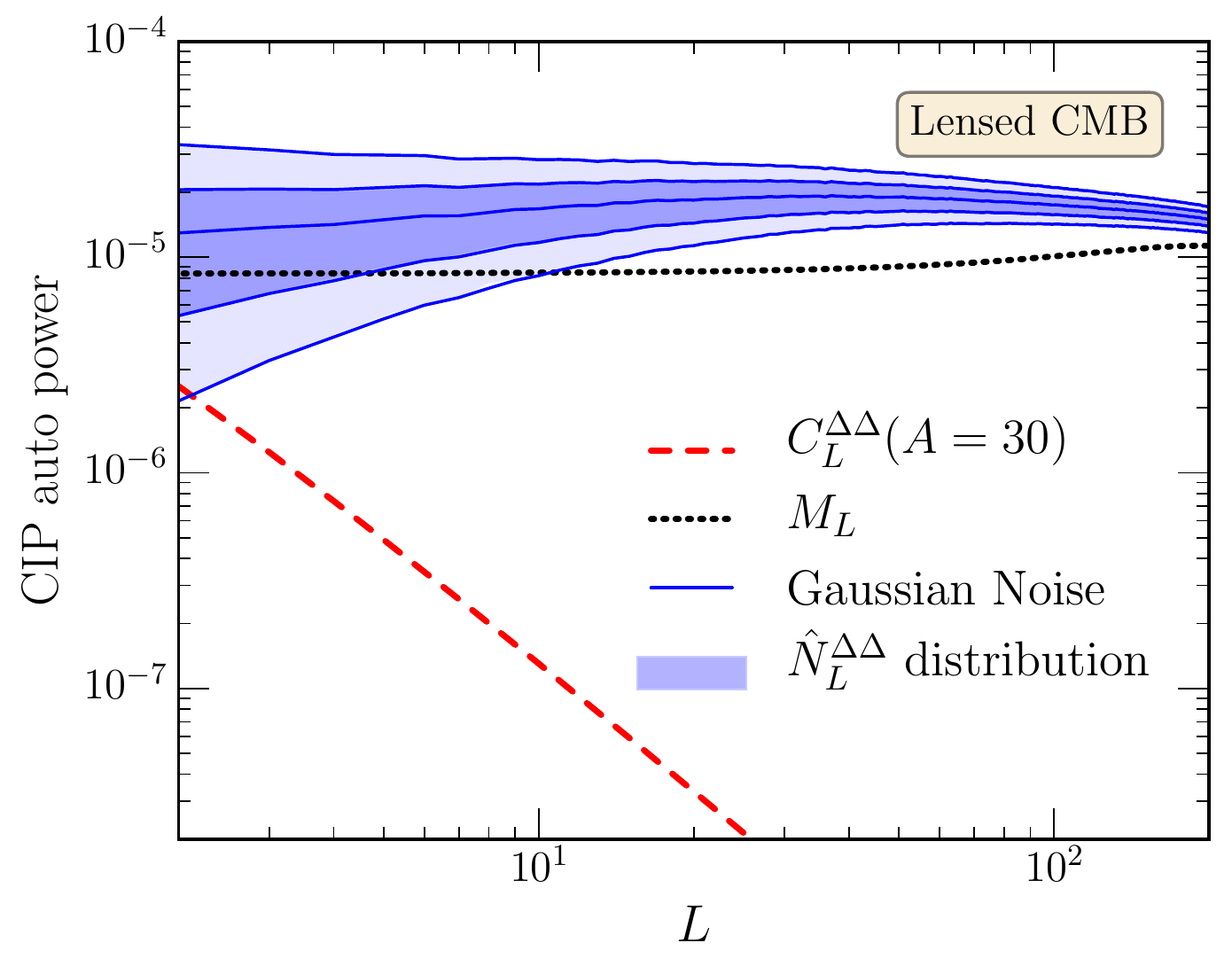}
          \caption{        
          CIP estimator noise power $\hat{N}_{L}^{\Delta\Delta}$ using lensed CMB maps.
          Shown are the mean, 68\% and 95\% confidence bands (shaded) of 4000 realizations of the estimator in
          the absence of a CIP signal.   The mean shows an excess of nearly a factor of two from the
          Gaussian CMB expectation 
          $\AL$ (dotted line) from Eq.~(\ref{eqn:normalization}).   The confidence bands still match the $\chi^2$ expectation of
          Gaussian noise with this enhanced mean (solid lines).  For reference we show a true CIP signal with $A=30$ (dashed line).
          }
           \label{fig:DDlensed}
\end{figure}

\begin{figure}
          \includegraphics[width=\columnwidth]{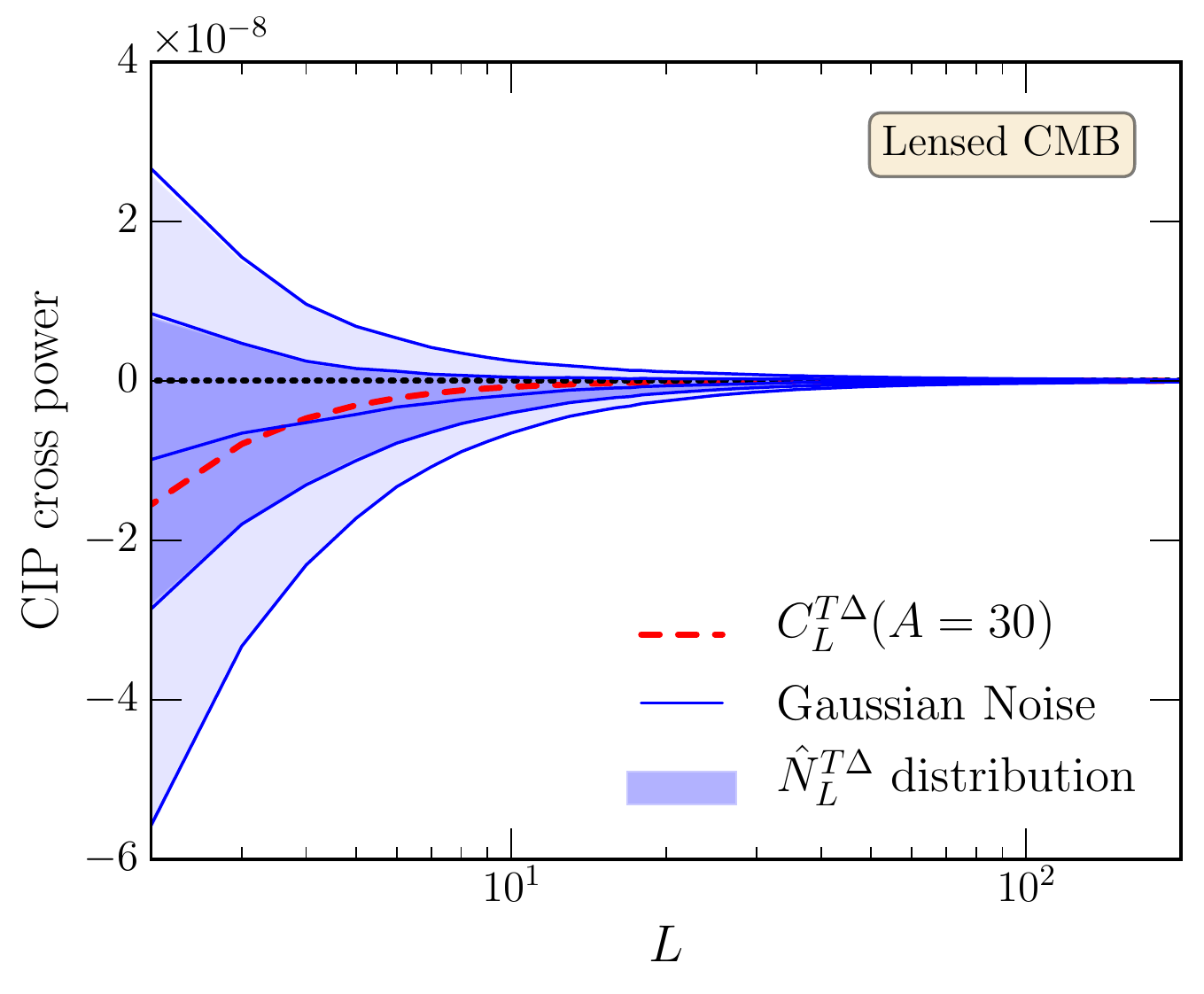}
          \caption{          CIP noise cross power $\hat{N}_{L}^{T\Delta}$ using lensed CMB maps.
          Shown are the mean, 68\% and 95\% confidence bands (shaded) of 4000 realizations of the estimator in
          the absence of a CIP signal.   The mean carries a nonzero contamination due to the lensing-ISW bispectrum \cite{Smith:2006ud,Lewis:2011fk,Kim:2013nea,Ade:2015dva} with the Gaussian expectation of zero shown for reference (dotted line).  	  The confidence bands still match the Wishart expectation of
          Gaussian noise given the mean power spectra (solid lines).     For reference we show a true correlated CIP signal with $A=30$ which has a similar spectrum to the contamination (dashed line).
           }  \label{fig:TDlensed}
\end{figure}

For each realization, we repeat the steps in Eq.~(\ref{eqn:filteredfields}), (\ref{eqn:estimator}), (\ref{eqn:noiserealization})
and (\ref{eqn:noiseaverage})
but with these properly lensed temperature maps.   In Fig.~\ref{fig:maplensed} (top) we
show an example realization of the noise in the CIP estimator.     In comparison to
Fig.~\ref{fig:mapgaussian}, the noise is notably larger even though the CMB temperature
field that it is constructed from has the same power spectrum $C_l^{TT}$ by construction.

This additional noise contribution is from the non-Gaussianity of the lensed CMB.
In fact, if the CIP estimator is averaged over the cosmic variance of the CMB at recombination with
a fixed realization of the lensing potential $\phi$, lensing produces a bias in the CIP
map itself
\begin{equation}
\langle \hat{\Delta}_{LM}  \rangle\big|_\phi \ne 0,
\end{equation}
 due to the statistical anisotropy that it creates.
In Fig.~\ref{fig:maplensed} (bottom), we separately perform such an average over 40 unlensed $\tilde{T}_{lm}$ CMB realizations lensed by the same potential.   Notice that the
fake CIP signal induced by the fixed $\phi$ has roughly the same amplitude and spectrum
as the Gaussian CMB contribution to the CIP noise in Fig.~\ref{fig:mapgaussian} (top).

Of course once  averaged over random realization of the lens potential as in our main $\nsims$ realizations, the bias
due to the statistical anisotropy of the lens  becomes a non-Gaussian CMB source of zero mean noise.   In Fig.~\ref{fig:DDlensed}, we show that the non-Gaussian lensing contribution nearly doubles
the noise power in $N_L^{\Delta\Delta}$.    Furthermore, the lensing-ISW correlation \cite{Smith:2006ud,Lewis:2011fk,Kim:2013nea,Ade:2015dva} produces
a nonzero $N_L^{T\Delta}$ as well (see Fig.~\ref{fig:TDlensed}).  This fake CIP-temperature
correlation has a spectrum that is similar but not identical to that predicted by the curvaton model and so
error propagation in its removal is important to model. 

In spite of their origin in the non-Gaussianity of the lensed CMB, these excess contributions
to the noise in the CIP estimator are nearly Gaussian distributed.
In Fig.~\ref{fig:DDlensed} and \ref{fig:TDlensed}, we compare the distributions of 
$\hat N_L^{\Delta\Delta}$ and $\hat N_L^{T\Delta}$ to the Gaussian noise expectation.
Even the tails of the distribution as displayed by the 95\% confidence region are well modeled
by a multivariate Gaussian.

\begin{figure}
\centering
          \includegraphics[width=\columnwidth]{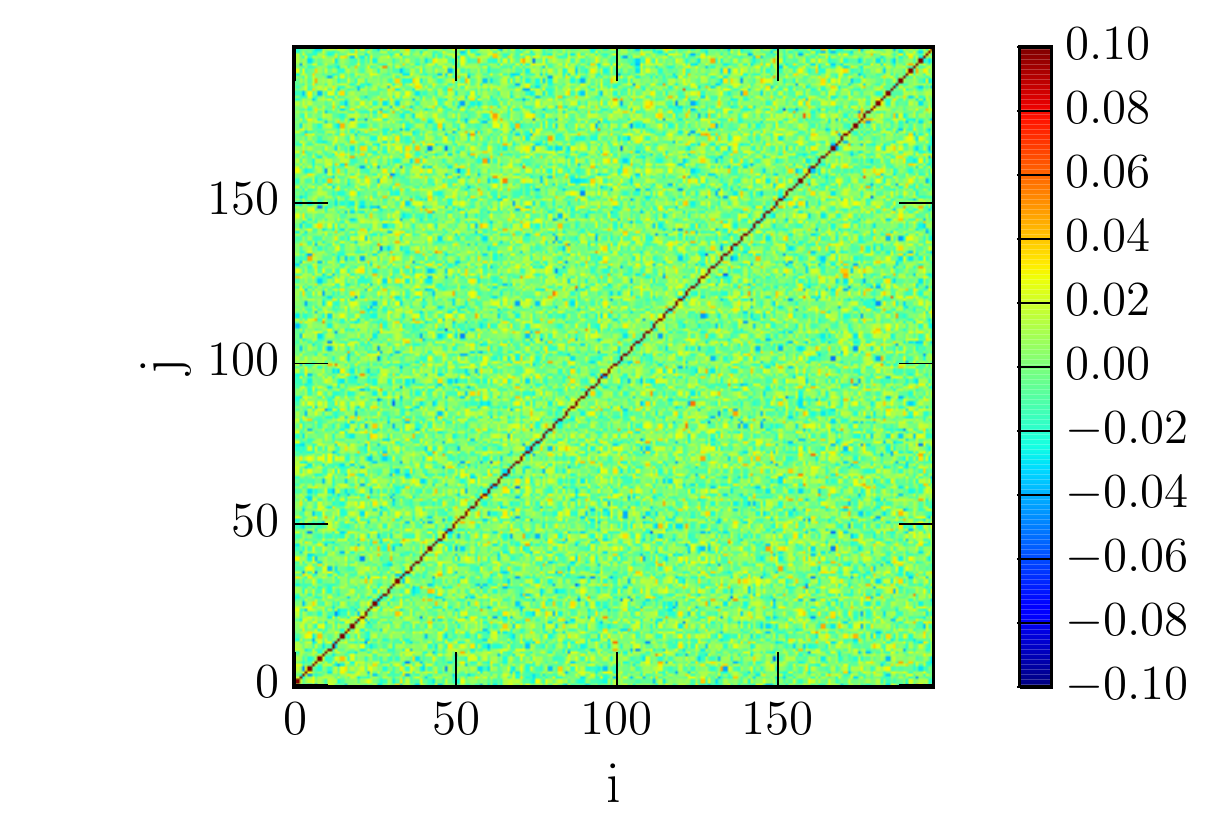}
         \caption{Correlation matrix $R_{ij}$ between multipoles $L_i$ and $L_j$ of the CIP noise power $\hat{N}_{L}^{\Delta\Delta}$ (Eq.~\ref{eq:Rij}) using 4000 realizations of the lensed CMB maps. The diagonal is removed to display the off-diagonal terms which fluctuate around
         a mean of $R = 0.00053$ with an r.m.s.~$\sigma_R=0.016$ as consistent with
         the finite number of realizations.}  
         \label{fig:RB}
\end{figure}

For the CIP estimator noise to be  Gaussian random, the covariance of the noise power
must be diagonal in $L$, not just Wishart distributed in its variance.To test this, we construct the covariance matrix of $\hat N_L^{\Delta\Delta}$:
\begin{equation}
C_{ij}  \equiv \langle \hat N_{L_i}^{\Delta\Delta} \hat N_{L_j}^{\Delta\Delta} \rangle -  N_{L_i}^{\Delta\Delta} N_{L_j}^{\Delta\Delta}.
\end{equation}
In Fig.~\ref{fig:RB}, we show the correlation matrix
\begin{equation}
R_{ij} = \frac{C_{ij}}{\sqrt{ C_{ii} C_{jj}}}. \label{eq:Rij}
\end{equation}
The off-diagonal $i\ne j$ elements are consistent with zero up to the expected $\nsims^{-1/2}$ statistical
fluctuations due to the finite sample. 

\begin{figure}
\centering
          \includegraphics[width=\columnwidth]{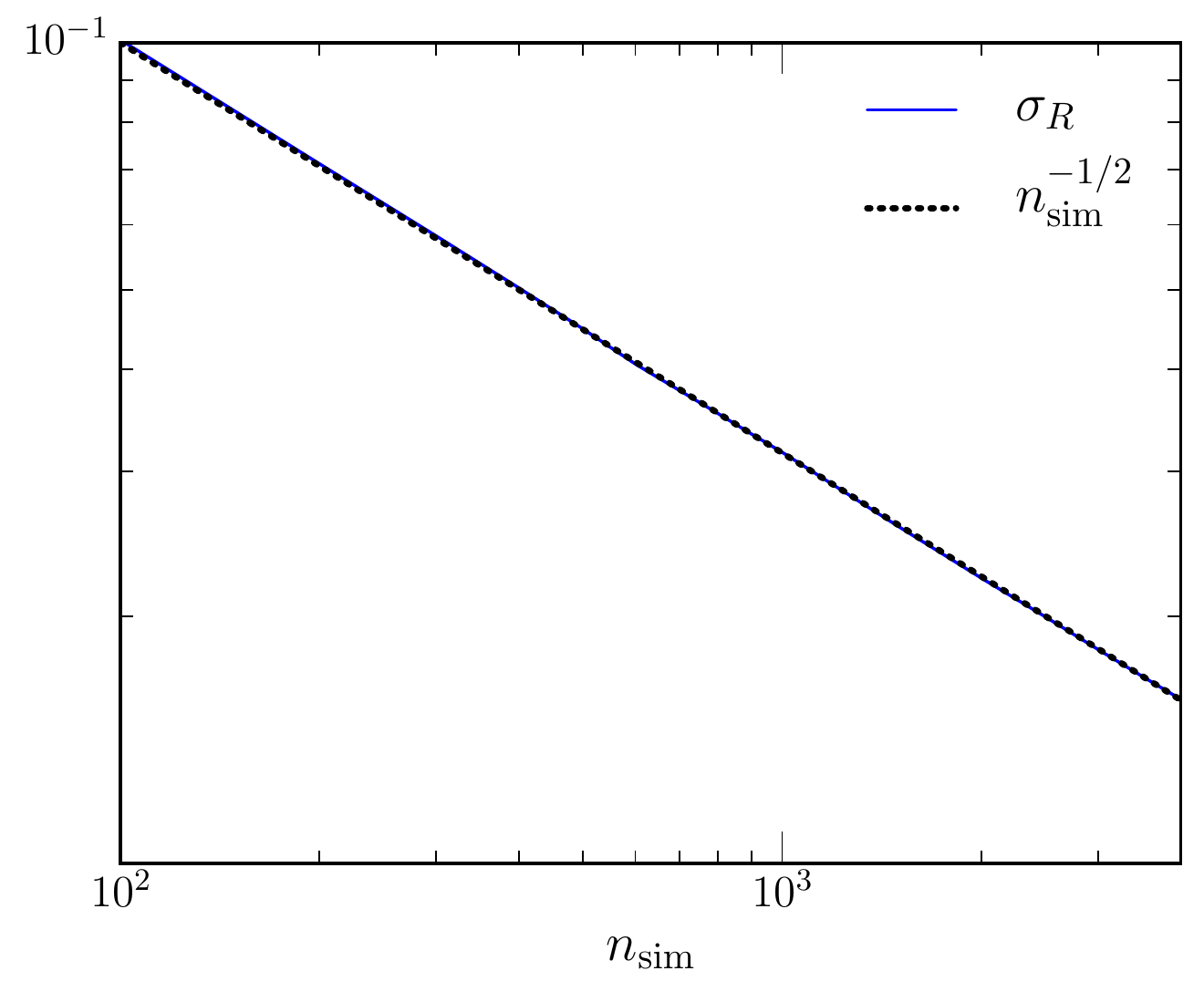}
         \caption{The r.m.s fluctuations in the off-diagonal correlations $\sigma_R$ vs.~the number of simulations $n_{\mathrm{sim}}$ for multipoles $L \in [2,200]$. The dispersion $\sigma_R$ (solid) scales as $n_{\mathrm{sim}}^{-1/2}$ (dotted), as expected for statistical fluctuations around an insignificant correlation. }
         \label{fig:RBrms}
\end{figure}

To quantify these bounds, we calculate the mean of the off-diagonal correlations
\begin{equation}
R = \frac{2}{n_L(n_L-1)} \sum_{i,j>i}   R_{ij} ,
\end{equation}
where $n_L$ is the number of multipoles $L \in [2,200]$,
and find a negligible $R=0.00053$ and the 
variance
\begin{equation}
\sigma_{R}^2 =\frac{2}{n_L(n_L-1)} \sum_{i,j>i}  R_{ij}^2 - R^2
\end{equation}
which gives $\sigma_R=0.016$.    
In Fig.~\ref{fig:RBrms}, we further test that the scaling of $\sigma_R$ with $\nsims$ by using the first $\nsims$ of the full 4000 set shows no sign of 
a deviation from $\nsims^{-1/2}$ that would indicate an underlying correlation. 

\begin{figure}
\centering
          \includegraphics[width=\columnwidth]{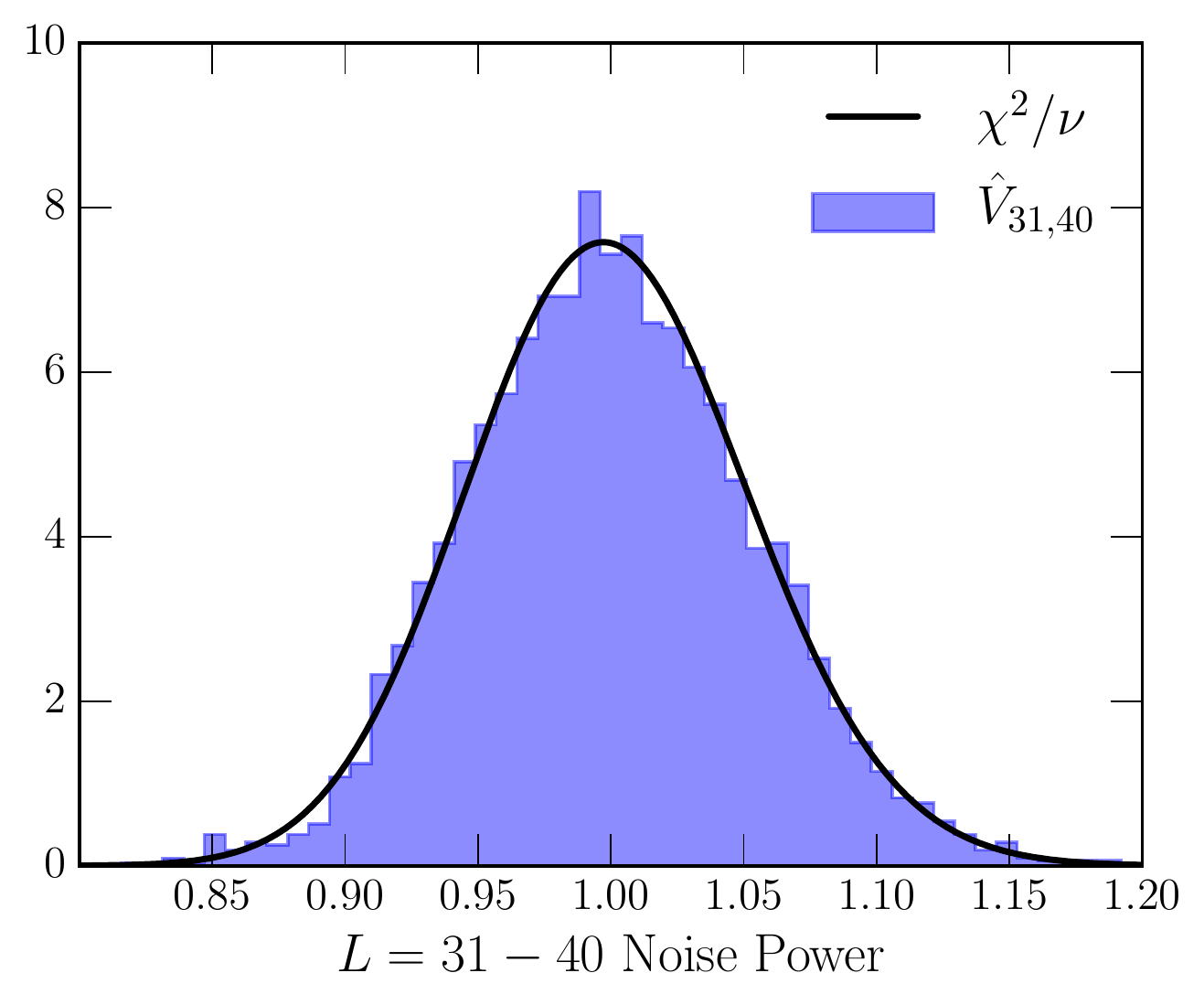}
         \caption{ Binned and normalized noise power $\hat V_{31,40}$ distribution for the
         multipole band  $L=31-40$.   The distribution is well modeled by the Gaussian noise expectation that $\nu \hat V_{31,40}$ is distributed as a $\chi^2$ with $\nu=720$ degrees of
freedom.}
         \label{fig:DDhist}
\end{figure}

Finally we test that binning the multipoles 
decreases the variance in the noise power in the manner expected for independent Gaussian random
variables.   The statistic 
\begin{equation}
\nu \hat V_{L_1,L_2}  \equiv \sum_{L=L_1}^{L_2} (2L+1) \frac{ \hat N_L^{\Delta\Delta}}{N_L^{\Delta\Delta}}
\end{equation}
should be distributed as a $\chi^2$ with $\nu$ degrees of freedom (see Eq.~\ref{eqn:chi2})
 where
\begin{equation}
\nu = \sum_{L=L_1}^{L_2} (2L+1) =(L_2+1)^2-L_1^2.
\label{eq:nu}
\end{equation}
In Fig.~\ref{fig:DDhist}, we show that the distribution 
of $\hat V_{31,40} $ in the 4000 simulations is in good agreement with the $\chi^2$ distribution. Other bins show the same level of agreement, which directly tests 
the approach taken in the next section of combining the information from each multipole
as if they were independent. 

\begin{table}[tbph]	
\caption{ $2\sigma_A$ detection threshold for $A$ given various noise assumptions for
CIPs that are correlated
with CMB temperature fluctuations and uncorrelated.  Here CIPs are reconstructed to multipole $L_{\rm CIP}=100$  from quadratic estimators with cosmic-variance limited CMB temperature
measurements out to $l=2500$.}
\label{tab:2sigma} 
\setlength{\tabcolsep}{0.5em}
\begin{ruledtabular}
\begin{center}
\begin{tabular}{lcc}
     & \multicolumn{2}{c}{$2\sigma_A$} \\ 
     Noise   & Corr. CIP & Uncorr. CIP \\ \hline     	\noalign{\smallskip}
Gaussian CMB Analytic  & 24 & 45         \\     
Gaussian CMB Sim. & 25 & 47        \\ 
Lensed CMB Sim.   & 31 & 58       \\ 
\end{tabular}
\end{center}
\end{ruledtabular}
\end{table}	

\section{Forecasts}
\label{sec:results}

In the previous section we have shown that the non-Gaussianity of the lensed CMB nearly doubles the noise power of
the CIP estimator for cosmic-variance limited temperature measurements out to  $l = 2500$.   
This has a substantial impact on the detectability of CIPs that we address in this section.

Even when the significant non-Gaussian effects of lensing are included, 
the total noise in the CIP estimator
 is nearly Gaussian distributed (see Sec.~\ref{sec:lensingsim}). 
   This means that we can treat the total noise like detector noise or cosmic variance in the data analysis.   To estimate the impact of the lensing contributions on the detection of
 the CIP amplitude $A$, we use the {Fisher matrix}. We focus on CIP reconstruction using cosmic-variance limited measurements of the CMB temperature field out to $l = 2500$ as in the simulations of the previous section. We
 also simplify the analysis by assuming that all cosmological parameters except for $A$ are fixed.

Under these assumptions the Fisher matrix has a single entry  whose inverse is the
estimate of the variance of measurements of $A$, $\sigma_A^2$.  It combines the information from the various CIP power spectra and 
multipole moments
\beq 
	\sigma_A^{-2}  = \sum_{L=2}^{L_{\rm CIP}}  \sum_{XY,X'Y'} 
	\frac{\partial C_L^{XY}}{\partial A}	\left({\bf C}^{-1}_{L}
	\right)_{XY,X'Y'} \frac{\partial C_L^{X'Y'}}{\partial A},
	\label{eq:fisher}
\eeq
where as before  $X,Y \in  \{T, \Delta\}$ and
 ${\bf C}_L$ is the covariance matrix 
\beq
		{\bf C}^{XY,X'Y'}_{L} =\frac{{ \tilde{C}_L^{XX'} \tilde{C}_{L}^{YY'} + \tilde{C}_L^{XY'} \tilde{C}_{L}^{X'Y}}}{2L+1} .
\eeq
We only include $L<L_{\rm CIP}=100$ in order to remain in the regime where the separate-universe assumption made in constructing the estimator is valid \cite{He:2015msa}.

The difference between this covariance and that implied by the Wishart distribution of the previous section is
that we now include the sample variance of a real CIP signal
\begin{eqnarray}
\tilde C_L^{\Delta\Delta} &=& C_L^{\Delta\Delta} + N_L^{\Delta\Delta}, \nonumber\\
\tilde C_L^{T\Delta} &=& C_L^{T\Delta} + N_L^{T\Delta}, \nonumber\\
\tilde C_L^{TT} &=&  C_L^{TT}= N_L^{TT}.
\end{eqnarray}
The sample variance depends on the assumed value of $A$ and so we take its value 
to be the detection threshold $A= 2 \sigma_A$ for the various simulated and analytic noise spectra 
$N_L^{XY}$ from the previous section (see Tab.~\ref{tab:2sigma}).

To establish the baseline, we start with the analytic form for the noise given a Gaussian CMB in the absence of CIPs and lensing
 as in Ref.~\cite{He:2015msa},
\bea
N_L^{\Delta\Delta} &=& M_L, \quad N_L^{T\Delta} = 0,
\eea
for which 2$\sigma_A = 24$ with correlated CIPs.  
Employing instead the Gaussian CMB simulations for the noise spectra we obtain a
consistent  $2\sigma_A = 25$.  
Finally with the  lensed CMB simulations 
the threshold is raised to $2\sigma_A = 31$, a factor of $\sim$1.3 higher compared to the case of analytical estimate of the noise for Gaussian CMB.

Given that these detection thresholds are larger than even the largest curvaton motivated value, it is
also interesting to consider the case of uncorrelated CIPs $C_L^{T\Delta}=0$ with the same
$C_L^{\Delta\Delta}$. We find that the lensing noise contributes here as well 
by raising the $2\sigma$ threshold a factor of $\sim 1.3$ from 45 to 58.

If we take $L_{\rm CIP}=200$, the detection thresholds are slightly lower in the correlated case but the relative degradation from lensing is similar.  In the uncorrelated case, the results are insensitive to the exact value of $L_{\rm CIP}$ since most of the signal-to-noise comes from large scales for a nearly scale invariant auto-spectrum. 
\section{Conclusions}
\label{sec:conclusions}
A large scale CIP field modulates the two point statistics of small scale CMB anisotropies,
much like gravitational lensing.   The quadratic reconstruction of the former from the latter
can therefore be contaminated by lensing.  We have shown, using simulations, that
  non-Gaussian modulation by lensing provides an additional contribution to the noise in the
CIP estimator that is comparable to that of the cosmic variance of the small scale CMB modes
themselves. Moreover, the lensing-ISW bispectrum {(present in the absence of real CIPs)} \cite{Goldberg:1999xm,Smith:2006ud,Lewis:2011fk,Kim:2013nea,Ade:2015dva} provides a false signal that
resembles the CIP-temperature cross correlation in the curvaton model; such contamination similarly afflicts estimators of the amplitude $f_{\rm NL}$ of primordial local-type non-Gaussianity \cite{Goldberg:1999xm,Smith:2006ud,Lewis:2011fk,Kim:2013nea,Ade:2015ava}.

Despite these non-Gaussian effects in the CMB, the resultant noise in the CIP estimator
is nearly Gaussian by virtue of the central limit theorem.   We find that the noise power
is uncorrelated to good approximation -- no more than  0.1\% correlations averaged over all 
multipole pairs  
with pair fluctuations that are consistent with our finite sample of simulations and less than $2\%$.
Furthermore, binning of multipoles reduces the variance 
of the noise power in the same manner as a Gaussian random field.

Treating the lensing contamination as excess Gaussian random noise, we  estimate
its impact on the detection of CIPs using the Fisher information matrix.   
 Assuming cosmic-variance limited measurements of the CMB temperature anisotropies out to $l = 2500$, we find that the detection thresholds for the CIP amplitude parameter $A$ are raised by a factor of 1.3 when the lensing bias is included for both the uncorrelated and correlated CIPs.   
 Here we have employed CIP reconstruction up to the separate-universe limit $L_{\rm CIP} = 100$ and assumed all other cosmological parameters are fixed.
 
 Our methodology, which employs direct simulation of lensing, is also straightforward to generalize to
 cases where measurement noise,  systematic effects and other cosmological parameters
 are included.   In these cases the lensing contamination is treated as an 
 additional signal that is jointly modeled along with the CIP contributions and prior information in a realistic
 data pipeline.  
 Cosmological parameter uncertainties can make the lensing contamination even more
 important given parameter degeneracies.   For example the 
 ISW-lensing contamination \cite{Smith:2006ud,Lewis:2011fk,Kim:2013nea,Ade:2015dva}, depends on the cosmic acceleration model and takes
 a similar form to the CIP cross correlation.   Fortunately, the lensing contamination to the
  CIP auto correlation takes a very different form which can be used to break degeneracies.
  
 Likewise, CMB polarization information can also assist in distinguishing lensing from
 CIPs as well as reduce the overall noise in the CIP estimator.    In particular quadratic reconstruction from the $E$ and $B$ modes has the highest signal-to-noise for lensing reconstruction \cite{Hu:2001kj}
 and the lowest for CIP reconstruction  \cite{He:2015msa}.   More generally, these additional fields
 provide consistency tests for the specific type of modulation expected by CIPs in the
 auto and cross power spectra of their quadratic estimators.
 
 Finally, there are consistency tests that are internal to just the temperature based
 CIP estimator.   The  estimator is constructed out of many pairs of temperature multipoles
 whose weights were chosen to be optimal in the absence of lensing.   Whereas the impact of CIP modulation does not depend strongly on the orientation of the modes, lensing does 
 since its effect vanishes if the deflection is in a direction orthogonal to the modes. 
 In principle, the pairs can be reweighted to de-emphasize the most contaminated modes
  at the expense of computational efficiency of the estimator.
  We leave these topics for future study.
  
 \begin{acknowledgments}
 We thank Cora Dvorkin and Duncan Hanson for stimulating discussions.  CH and WH were supported by U.S.~Dept.\ of Energy contract DE-FG02-13ER41958 and NASA ATP NNX15AK22G. DG is funded at the University of Chicago by a National Science Foundation Astronomy and Astrophysics Postdoctoral Fellowship under Award  AST-1302856. This work was supported in part by the Kavli Institute for Cosmological
 Physics at the University of Chicago through grant NSF PHY-1125897 and
an endowment from the Kavli Foundation and its founder Fred Kavli.  Computing resources were provided by the University of Chicago Research Computing Center.
\end{acknowledgments}

\appendix

\bibliography{chen_spires}
\end{document}